Assessing Bias and Precision in State Policy Evaluations: A Comparative Analysis of Time-Varying Estimators Using Policy Simulations

**Author list:** Max Griswold,[1] Beth Ann Griffin,[1] Max Rubinstein,[2] Mincen Liu,[3] Megan Schuler,[1] Elizabeth Stone,[4] Pedro Nascimento de Lima,[5] Bradley D. Stein,[2] and Elizabeth A. Stuart[6]

**Affiliations:**
[1]RAND, Washington, DC
[2]RAND, Pittsburgh, PA
[3]McGill University, School of Population and Global Health
[4]Rutgers University
[5]RAND, Santa Monica
[6]Johns Hopkins Bloomberg School of Public Health

**Corresponding Author:**
Max Griswold
1200 S. Hayes Street
Arlington, VA 22202
703-413-1100
griswold@rand.org

**Running title:** Assessing Bias and Precision in Evaluations

**Conflict of Interest:** The authors report no conflicts of interest.

**Sources of Funding:** This research was supported by a grant from the National Institutes of Health P50DA046351.

**Data and Computing Code Availability**: Data and code for this study are available at: https://github.com/maxgriswold/optic-time-vary

**Acknowledgements:** We thank participants at the 2024 Addiction Health Services Research Conference for their helpful feedback and comments on earlier drafts of this report.

**Abstract**

**Background** Evaluating state-level policies using panel data often assumes static treatment effects, yet many policies have effects that vary depending on time since implementation. This study compares the performance of panel data estimators in estimating time-varying policy effects to inform optimal methodological choices for policy evaluations.

**Methods** Using state-level opioid overdose mortality data from 1999-2016, we simulated four time-varying treatment scenarios, which correspond to real-world policy dynamics (ramp up, ramp down, temporary and inconsistent). We then evaluated seven commonly used policy evaluation methods: two-way fixed effects event study, debiased autoregressive model, augmented synthetic control, difference-in-differences with staggered adoption, event study with heterogeneous treatment, two-stage differences-in-differences and differences-in-differences imputation. Statistical performance was assessed by comparing bias, standard errors, coverage, and root mean squared error over 1,000 simulations.

**Results** Our findings indicate that estimator performance varied across policy scenarios. In settings where policy effectiveness diminished over time, synthetic control methods recovered effects with lower bias and higher variance. Difference-in-difference approaches, while offering reasonable coverage under some scenarios, struggled when effects were non-monotonic. Autoregressive methods, although demonstrating lower variability, underestimated uncertainty. Overall, a clear bias-variance tradeoff emerged, underscoring that no single method uniformly excelled across scenarios.

**Conclusions** This study highlights the importance of tailoring the choice of estimator to the expected trajectory of policy effects. In dynamic time-varying settings, particularly when a policy has an anticipated diminishing impact, methods like augmented synthetic controls may offer advantages despite reduced precision. Researchers should carefully consider these tradeoffs to ensure robust and credible state-policy evaluations.

# INTRODUCTION

There is wide interest in determining the effectiveness of state policies on health outcomes using observational panel data. In practice, policies may have effects that vary depending on the time since implementation. For example, pharmacy-access laws are effective at increasing naloxone distribution, and that the effect grows over time following policy implementation.[1] In other cases, an initial immediate policy effect may dissipate over time, such as school immunization policies, which may become less effective over time due to increased exemptions;[2] state tobacco taxes, which may have a smaller impact as real prices increase and the remaining population of smokers become more resistant to behavior change;[3] prescription drug monitoring programs, whose effectiveness may wane if patients adapt to use illicit sources of nonprescription opioids;[4] and state naloxone co-prescribing mandates, which can led to initially higher co-dispensing rates which subsequently decreased afterwards.[5] It is critical for researchers to be able to model policy effectiveness across time to help inform policymaker decision making.

Recently, there has been a large increase in the number of panel data estimators available to study the time-varying effects of policies, including new extensions of difference-in-differences (DID) event study approaches[6] and synthetic controls.[7] These methods differ in how treated states are compared to controls and in how they aggregate the data to estimate policy effects over time since implementation. In principle, if the assumptions of these approaches are met, each should produce unbiased estimates of policy effects across time since implementation, although the actual estimands of interest may differ across methods. We therefor expect that these estimators may differ in their statistical performance when estimating a time-varying effect *for a given treatment period*. As such, it is critical to evaluate the performance of these estimators in terms of the dynamics of policy effects following implementation.

This study explores the statistical performance of seven different panel data methods in estimating time-varying policy effects across different dynamic treatment effect scenarios. Building upon a simulation framework described in previous articles,[8–10] we compare the bias and precision of panel-data methods used to estimate dynamic time-varying policy effects. We examine methods across four scenarios intended to reflect a range of policy scenarios practitioners may encounter in applied work, specifically cases where the policy's impact (1) increases over time until a full effect is achieved (ramp up), (2) has an immediate effect that then dissipates overtime (ramp down), (3) increases before then decreasing to no effectiveness (temporary), and (4) varies over time since implementation before stabilizing (inconsistent).

# METHODS

In this section, we introduce the methods under comparison, the simulation set-up and procedure, and the evaluation criteria used to analyze methods. The study was approved by the corresponding author's Institutional Review Board with a waiver of consent.

## Causal inference notation and primary policy effects of interest

Using potential outcomes notation, let $Y_{it(1)}$ denote the potential outcome for state $i$ ($i = 1,...,50$) if the policy was in effect at time $t$, while $Y_{it(0)}$ denotes the potential outcome for state $i$ if the policy was not in effect at time $t$. Thus, each state has two potential outcomes at each relative treatment time point, representing the mortality rates that would be observed with and without the policy in effect. Our primary policy effect of interest is $[E[Y_{t=j,(1)} - Y_{t=j,(0)}]$, averaging across both states and relative treatment times in the j post policy periods. Let $A_{i,t=j} = \{0,1\}$ denote an indicator for whether state $i$ had the policy in effect at relative treatment time $j$. Then, $Y_{i,t}^{obs} = Y_{i,t=j(1)} * A_{i,t=j} + Y_{i,t=j(0)} * (1 - A_{i,t=j})$ denotes the observed outcome for state $i$ at relative treatment time $j$. Let $X_{it}$ denote the observed time-varying covariate of unemployment rate at the state level, which was included in all models discussed below.

**Empirical Methods Considered**

We compared results from seven statistical models applied to the simulated datasets: (1) a two-way fixed effect DID estimator using an event study specification (DID-ES); (2) a debiased autoregressive model (AR-DB); (3) an augmented synthetic control model with staggered adoption (ASCM);[11] (4) a staggered adoption difference-in-differences (DID-SA) estimator;[12] (5) a difference-in-differences event study with heterogenous treatment (DID-HT);[13] (6) a two-stage difference-in-differences estimator (DID-2S);[14] and (7) a difference-in-differences imputation estimator (DID-IMP).[15] We briefly describe each model and how they estimate time-varying policy effects in the following section. We chose these models based on previous policy simulation studies and on recently proposed design-based methods for addressing time-varying effect heterogeneity in policy evaluations.[6,9,16]

Two-way Fixed Effects Event Study (DID-ES). Event study models are commonly used to estimate time-varying policy effects using longitudinal data.[17] This approach estimates time-varying policy effects using a binary indicator corresponding to the number of periods since treatment was implemented. This model aims to control for confounding using unit and time fixed effects, along with covariate adjustment. For this approach, we estimated policy effects using the following specification:

$$Y_{i,t}^{obs} = \sum_{j=-m}^{n} \gamma_j \cdot D_{i,t-j} + \alpha_i + \delta_t + \beta \cdot X_{i,t} + \epsilon_{i,t} \qquad (1)$$

where $\gamma_j$ is an estimated effect for event-time $j$. The event-time indicates a period relative to the first year of treatment. $D_{i,t-j}$ is an indicator variable for treatment status in event-time, which equals one if treatment occurred within exactly $j$ periods since year t. Here, $-m, n$ denotes start and endpoints in event-time, which can differ depending on when a state $i$ was first treated.

Throughout this paper, we only display results for j = {1, 2, 3, 4, 5}, though we estimated effects for all anticipation periods (i.e., $j < 0$), except j = -1 which was an omitted time-period, and additional treatment event-periods (i.e., $j > 5$). The specification also includes unit and time fixed effects ($\alpha_i$ and $\delta_t$, respectively). Like a DID estimator, the fixed effects remove time-invariant confounding between treated and untreated units. Standard errors from this model were estimated by applying a state-level cluster adjustment using the sandwich package in R.

Debiased Autoregressive Model (AR-DB). We employed a debiased autoregressive (AR-DB) model to estimate the causal effects of policy interventions while addressing biases introduced by lagged outcome adjustments.[18] Previous simulation studies demonstrated that debiased autoregressive models perform well when estimating state policy effects.[8–10] These models include functions of one or more lagged measures of the outcome (e.g., $Y_{i,t-k}$) as covariates to control for potential confounding due to differences in prior outcome levels and improve estimation of policy effects when outcomes are highly autocorrelated, as with annual measures of state-level opioid-related mortality. However, standard autoregressive models may induce bias because of prior treatments on the lagged outcomes. By contrast, the debiased AR model adjusts for estimates of the prior outcome absent treatment, thereby removing the biasing effects of the treatment on the previous outcome and enabling more accurate inference.

The model we estimate posits that $Y_{i,t}$ is a function of the outcomes absent treatment for $k$ prior time periods $Y_{i,t-1}, \dots, Y_{i,t-k}(0)$, exogeneous time-varying covariates $X_{i,t}$, and the history of treatment up to l prior time periods $A_{it}, \dots, A_{it-l}$. Additionally, we posit that $Y_{i,t-k}(0) = Y_{i,t-k} - \sum_{z=0}^{l} \theta_z A_{i,t-k-z}$, thus giving the model,

$$Y_{i,t} = \alpha_i + \beta \cdot X_{i,t} + \sum_{b=1}^{k} \delta_b \left( Y_{i,t-b} - \sum_{z=0}^{l} \theta_z A_{i,t-b-z} \right) + \sum_{z=0}^{l} \theta_z A_{i,t-z} + \epsilon_{i,t}, \qquad (2)$$

where $\epsilon_{i,t}$ is a conditionally mean-zero error term. We estimate the model using the NLS function in the stats 4.2.2 R package.

Augmented Synthetic Control Model (ASCM). We also considered a partially pooled augmented synthetic control model (ASCM) for the staggered adoption setting.[7,11] ASCM is closely related to the synthetic control method, which uses untreated units to create a weighted synthetic control group that closely matches a single treated unit in pre-treatment outcomes and covariates. ASCM for staggered adoption extends the synthetic control method to settings with multiple treated units and improves the weighting of control units by adding a bias-correction procedure.

This approach uses two steps to estimate time-varying policy effects. First, ASCM estimates separate synthetic control unit weights for each treated unit by using a partial pooling approach to minimize imbalance between pre-treatment outcomes in treated units and the weighted average control unit. Second, using this weighted average control unit and additional adjustment covariates, ASCM solves an optimization problem to estimate a synthetic counterfactual outcome under control for each treated unit. This counterfactual outcome is used to calculate time-varying policy effects for a given number of post-treatment periods. We used the augsynth v0.2 R package to implement ASCM.

Difference-in-Difference with Staggered Adoption (DID-SA). We examined estimation of time-varying policy effects using an approach to DID analysis suggested by Callaway and Sant'Anna.[12] This approach comprises two steps for estimating an average time-varying policy effects across years since enactment. First, it estimates cohort-specific time-varying policy effects; it then aggregates those cohort average policy effects to estimate the average time-varying policy effect across the sample.

This estimator uses a combination of probability weighting and an outcome regression in step 1 to construct a consistent, doubly robust estimate of the time-varying policy effect, when either the probability weighting model or outcome regression model is correctly specified. Cohort-specific average policy effects correspond to estimates of the time-varying policy effect for units that received treatment at the same time. For step 2, the estimator weights estimated group-time average policy effects into a final set of time-varying effects across the sample. We estimated this model following procedures in Callaway and Sant'Anna, 2021 and used the DID v2.1.2. package in R to estimate all DID-SA models.

Event Study with Heterogenous Treatment (DID-HT). Sun and Abraham provide an alternative two-step strategy to estimate time-varying policy effects.[13] In the first step, an event-study design is modified to use cohort-specific interaction terms rather than event-time interaction terms. Then, the estimated cohort-specific interaction estimates are weighted by the relative share of each cohort within an event-time period, then aggregated to construct time-varying policy effects.

$$Y_{i,t}^{obs} = \sum_{e} \sum_{j=-m}^{n} \gamma_{e,j} \cdot 1(E_i = e) D_{i,t-j} + \alpha_i + \delta_t + \beta \cdot X_{i,t} + \epsilon_{i,t} \tag{3}$$

We estimated Sun and Abraham event study models using the Fixest v0.11.1 package in R, using a cohort-interaction-weighted estimator:

Two-Stage DID. We estimated an additional set of models using the two-stage DID (DID-2S) design adapted to event-studies, as proposed by Gardner.[14] This approach uses two steps to estimate time-varying policy effects. In the first stage, the method identifies varying heterogenous effects across groups g and time periods t using untreated units; the second stage removes these effects to correctly estimate average time-varying policy effects for n periods.

Stage One:

$$Y_{i,t}^{Never-treated} = \lambda_g + \delta_t + \epsilon_{i,t} \tag{4}$$

Stage Two:

$$Y_{g,t} - \widehat{\lambda_g} - \widehat{\delta_t} = \sum_{j}^{n} \gamma_{g,t,j} D_{g,t,j} + \beta \cdot X_{i,t} \tag{5}$$

We used the did2s v1.0.2 package in R to estimate these models.

DID Imputation Estimator (DID-IMP). The last model examined was a DID estimator based on imputation.[15] This estimator starts by first predicting ("imputing") a counterfactual policy outcome within untreated groups. Then for each treated unit and time, the counterfactual outcome is subtracted from the observed outcome to obtain unit-time-specific policy effect estimates. Lastly, the unit-time policy effects are averaged by event-time to construct average event-time policy effects.

Stage One:[15]

$$Y_{i,t}^{Never-treated} = \alpha_i + \delta_t + \beta \cdot X_{i,t} + \epsilon_{i,t} \tag{6}$$

Stage Two:
$$\gamma_{i,t} = Y_{i,t} - \widehat{\alpha_t} - \widehat{\delta_t} - \hat{\beta} \cdot X_{i,t} \tag{7}$$

We used the didimputation v0.3.0 package in R to estimate these models.

**Simulation Design**
Since it is impossible to test model assumptions in practice, we used a simulation study with a known data-generating process to assess the relative performance of these statistical models under several different scenarios related to the potential time-varying effects of a state policy. Our current work builds on prior simulations that compared the performance of statistical methods in estimating policy effects on total firearm deaths and opioid-related mortality.[8–10]

Observed Data. We used 1999- 2016 data on fatal overdoses obtained from the National Vital Statistics System – Multiple Cause of Death mortality files for our outcomes. Opioid-related overdose deaths were identified using ICD-10 external cause of injury codes X40-X44, X60-64, X85, and Y10-14. We aggregated injury codes and calculated state-level overdose rates per 100,000 state residents. We included state-level unemployment rate per 100 state residents as a covariate in all model specifications [19] to potentially increase model precision. The policy variable represents a hypothetical policy intended to reduce opioid-related mortality and is simulated to reflect staggered policy adoption.

Simulating Time-Varying Policy effects. We simulated treatment status by state and year so we could assess relative model performance under a known data-generating process. To do so, for each state and year, we generated a binary-indicator $T_{i,t}$ indicating whether a hypothetical state policy was in effect at some point between 1999 and 2016. Once a policy was in effect, the state remained treated for subsequent years. Control states were coded as $T_{i,t} = 0$ for the entire study period. For each simulation dataset, we randomly selected a given number of treated states (5, 25, or 45 states in each run), and applied treatment to a randomly selected year. We did not incorporate any additional confounding in order to allow us to carefully examine the relative performance of the models without this added layer of complexity.

We then generated synthetic outcomes for treated states based on four time-varying effect scenarios. For control units, we set the outcome variable to be the same as the observed state-specific, year-specific opioid overdose rate. For treated units, we set the outcome value to be equal to the observed opioid overdose rate ($Y_{i,t}$) for state $i$, year $t$, and augmented by a simulated policy effect ($\alpha_k$) for $k$ years since policy implementation:

$$Y^*_{i,t} = Y_{i,t} + \sum_{k=1}^{5} \alpha_k$$

We chose values of simulated policy effects to be small, within a 0-10% change from the annual opioid-related mortality rate across states, corresponding to standardized mean (effect size) differences between 0 and 0.2.[20] Note that while effects vary post-treatment across scenarios, the average policy effect following treatment in each scenario is the same (a 5% decrease in the observed overdose rate) and reflects the size of effects found in prior studies of opioid state policies and used in simulation studies.[5,8,9,21–23]

We utilized time-varying scenarios (Figure 1) meant to represent policy effects analysts could reasonably encounter in applied work, specifically: (i) a policy that takes five years to (linearly) increase until a full effect is achieved (**Ramp Up**); (ii) a policy that has an immediate and large effect that dissipates linearly over time (**Ramp Down**) (iii) a policy that increases in effectiveness before then decreasing to no effectiveness (**Temporary**); and (iv) a policy with effects that vary over time, increasing in effectiveness initially, followed by a decrease in effectiveness before stabilizing (**Inconsistent**).

**Figure 1: Policy effects by simulation scenario**

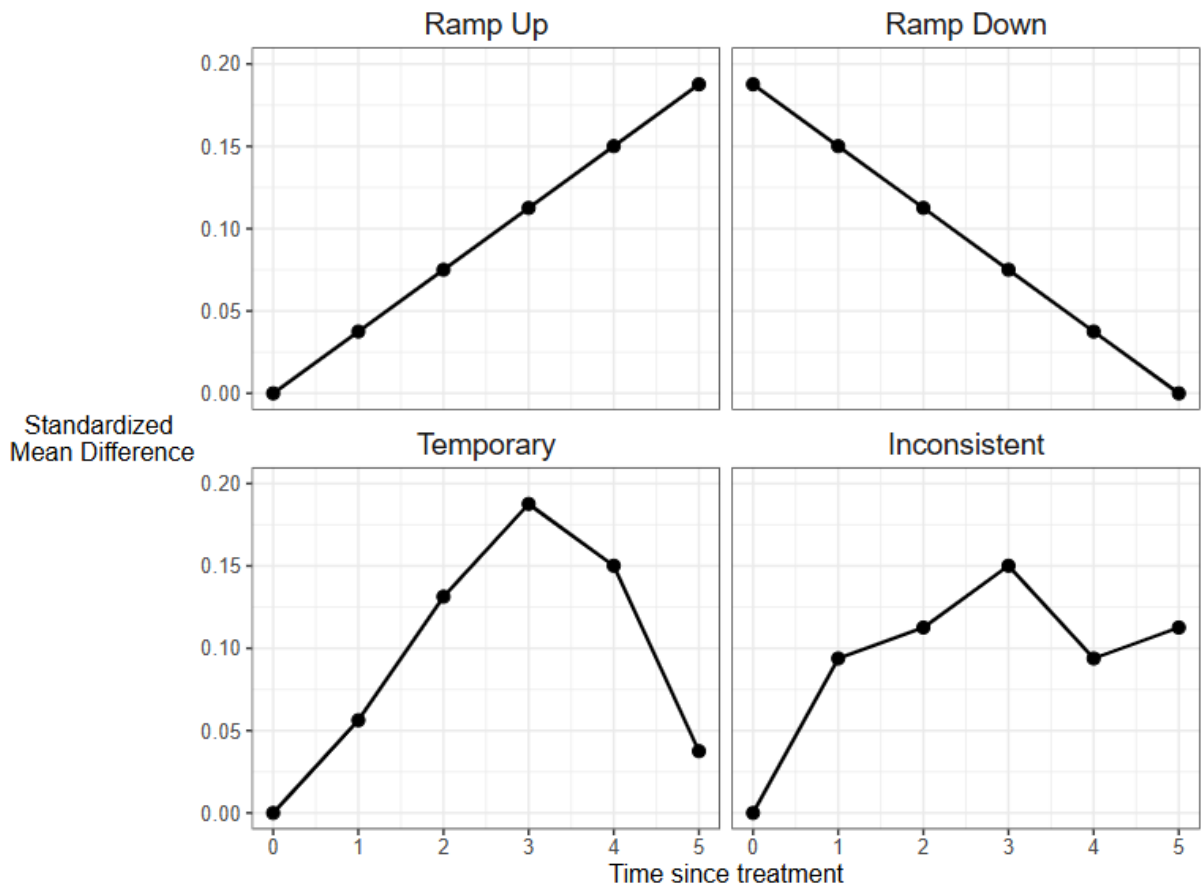

We generated 1000 simulated datasets for each scenario, which varied in the number of treated units (5, 25, or 45 treated states) and the four effect scenarios described above.

Performance Metrics. For each simulation scenario, we calculated performance metrics for each time-varying effect $\alpha_t$ for j belongs to [1, 2, 3, 4, 5]: average standardized bias and variance, root mean squared error, and coverage. More formally, we examined:

(1) *Average absolute bias.* For each method and time point j, we calculated the average absolute difference across simulations between the estimated effect and the true effect:
$Bias_{\alpha_t} = \sum_{k=1}^{1000} \frac{|\hat{\alpha}_{t,k} - \alpha_t|}{1000}$

*(2) Average standard error of estimated effects across simulations.* We summarized the extent of variation in estimated effects by calculating the standard error across simulation runs: $\overline{sd_{\hat{\alpha}_t}} = \sqrt{\sum_{k=1}^{1000} \frac{(\hat{\alpha}_{t,k} - \bar{\alpha}_t)^2}{1000}}$

(3) *Coverage:* We calculated the percentage of iterations in which the 95% confidence interval of the estimated event-time effect contained the true effect.

(4) *Root Mean Squared Error (RMSE).* Lastly, we calculated RMSE to quantify the performance of the modeling procedures considering bias and variance together, calculated as $RMSE_{\alpha_t} = \sqrt{\sum_{k=1}^{1000} \frac{(\hat{\alpha}_{t,k} - \alpha_t)^2}{1000}}$

All simulations were conducted in the OPTIC v1.0.1. package in R. Code for simulation reproduction, as well as the empirical data is available at https://github.com/maxgriswold/optic-time-vary.

**RESULTS**

Below we discuss how bias, variance, coverage, and RMSE of each method vary across time-varying treatment scenarios, highlighting performance with 25 treated units (findings for other numbers of treated states are similar in overall trends and are in Supplemental Digital Content Section 1).

**Examining average absolute bias of estimated effects across simulations**
Figure 2 shows the average absolute bias of estimated effects for each method across our four time-varying scenarios. Across all scenarios and years, ASCM has the lowest average standardized absolute bias (0.1) while AR-DB has the highest (0.13) Performance of the DID methods (DID-ES, DID-SA, DID-HT, DID-2S, DID-IMP) was highly similar with : DID-ES having the lowest average absolute bias (0.117) among these methods.

**Figure 2: Median standardized bias across simulation iterations**

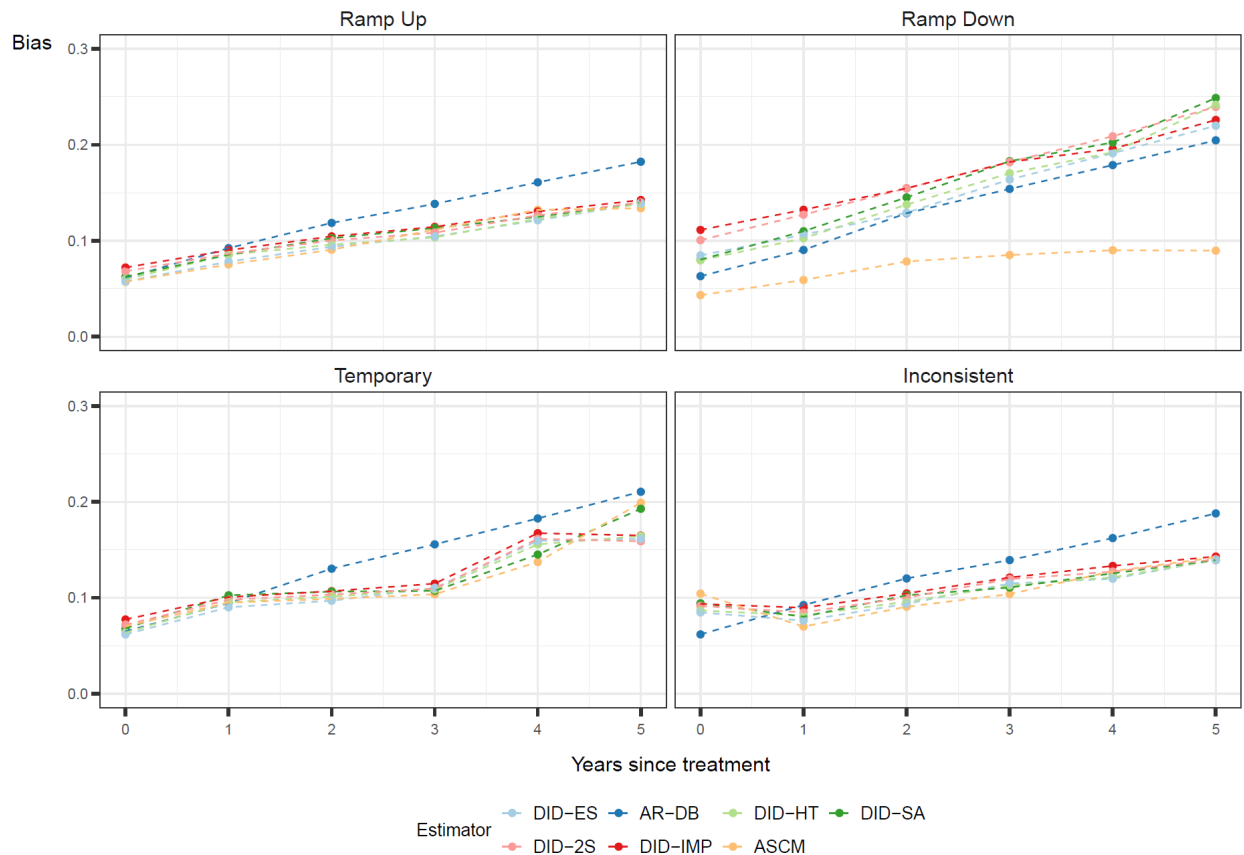

We see some interesting trends across the time-varying scenarios. First, the methods tend to have the lowest absolute bias in the **Ramp Up** scenario, with an average of 0.106 across methods and time periods, and the highest absolute bias in the **Ramp Down** scenario, with an average of 0.144. Within each scenario, DID-ES has the lowest absolute bias in **Ramp Up** (0.099), **Temporary** (0.113), and **Inconsistent** (0.104) and ASCM in **Ramp Down** (0.074). The DID methods all have poor performance in the **Ramp Down** scenario, with the estimated absolute bias increasing as the true effect size decreases (Figure 2).

We also investigated how an increased sample size might lead to changes in rank-ordering of method performance by generating synthetic states (Supplemental Digital Content Section 2). We generally find that results are similar to those in Figure 2, though bias tends to flatten across years-since-treatment periods. However, with additional sample size, ASCM exhibits vastly reduced bias compared to all other methods across scenarios, indicating this approach disproportionate benefits from an increased sample size.

In general, the DID based methods tend to increase in bias as years-since-treatment increase and ASCM tends to have the lowest bias across scenarios and years-since-treatment periods.

**Examining standard error of estimated effects across simulations**

Across scenarios and periods, the AR-DB method tends to have effects estimated with the least variability (Figure 3), yielding the lowest reported standard error of 0.742, followed by DID-IMP with 0.919. ASCM produces estimated effects that have the greatest variability with a standard error of 1.12. In general, we find similar results across DID estimators, with DID-IMP having the lowest standard error, compared to DID-2S, which consistently has the highest standard error across simulation runs (1.11). The magnitude of difference between the methods with the lowest standard error – AR-DB and DID-IMP – is marginal within most scenarios, except for the **Ramp Down** scenario.

**Figure 3: Median Standard error of estimated effects across simulation iterations**

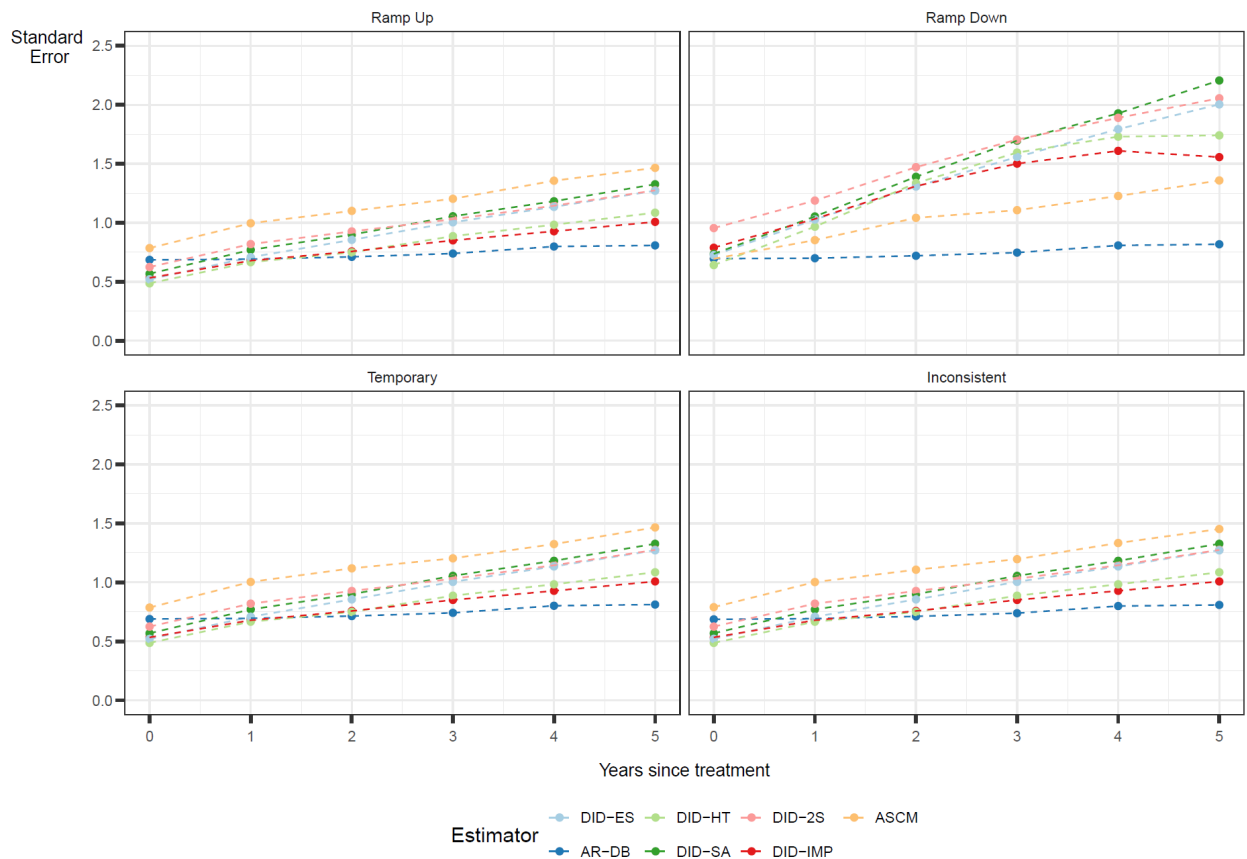

Generally, estimators have increasing standard errors as years-since-treatment increase, except for the AR-DB which has consistently small standard error across most periods. Further, the increase in standard error by period are of relatively similar magnitude for each method across scenarios besides **Ramp Down**. Notably in **Ramp Down**, ASCM has a much lower standard error than DID estimators compared to other scenarios (e.g., 1.04 vs 1.3 for DID-IMP), while the standard error of ASCM is larger than these methods in all other scenarios.

**Examining simulation coverage**

We next present model coverage in Figure 4 noting immediately that virtually none have values that are close to the 95% level desired, except for some models in **Ramp up**. Across scenarios and time periods, AR-DB has the lowest average coverage (75.6%), while ASCM has the highest average coverage (98.9%). More specifically, ASCM average coverage is close to one in most scenarios (99.2%, 99.6%, 97.9%, 98.8% in **Ramp up**, **Ramp down**, **Temporary**, and **Inconsistent**, respectively) suggesting anti-conservative coverage rates, while AR-DB is more variable but generally with coverage rates that are too low (77.1%, 74.2%, 74.4%, 76.7%). For ASCM, the wide coverage is driven by the large model based standard errors and for AR, the low coverage is due to extremely small model based standard errors (see Supplemental Digital Content Section 3).

**Figure 4. Coverage across simulation iterations**

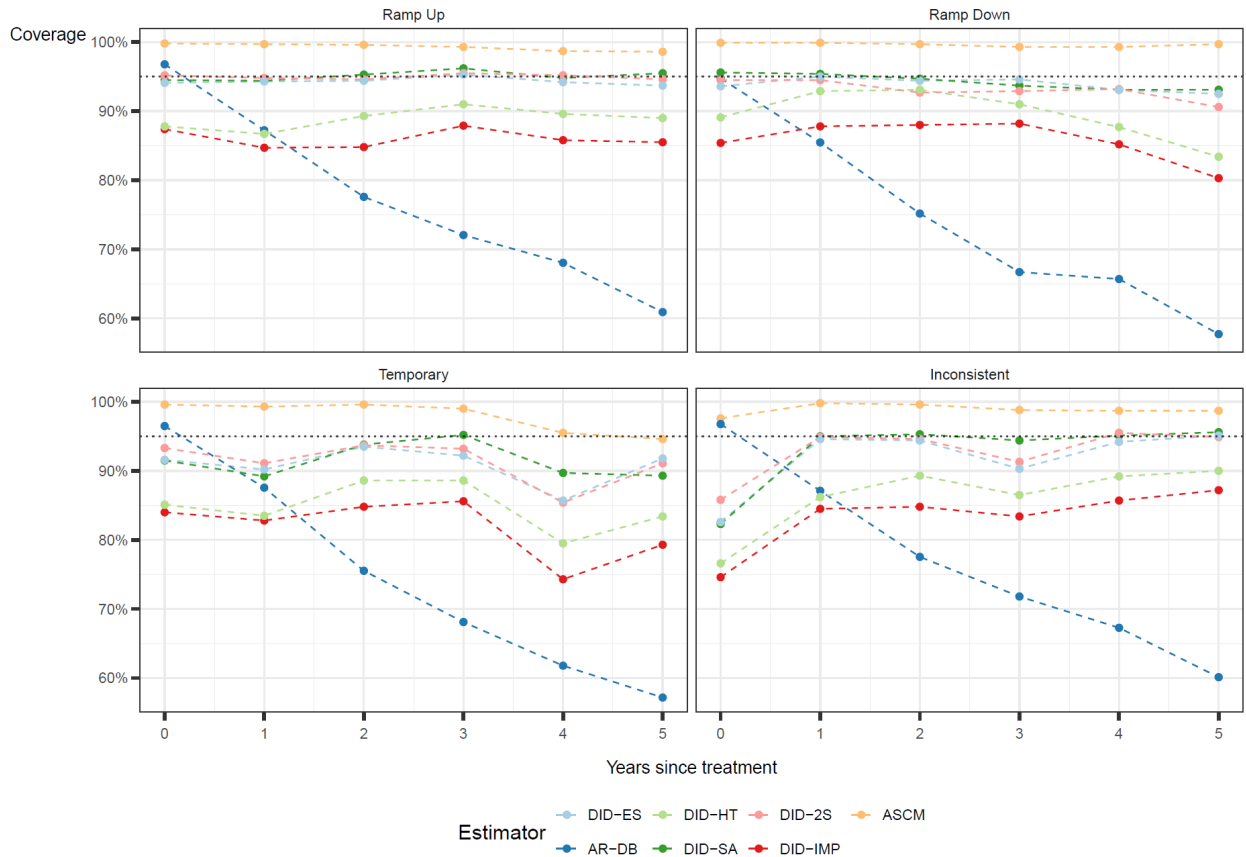

Among DID methods, DID-IMP exhibits the lowest average coverage (84.3%), while DID-SA has the highest coverage (93.4%). The rank-order of DID models by coverage are similar across scenarios, with DID-SA exhibiting the highest average coverage and DID-IMP having the lowest average coverage. DID methods also appear to have coverage in a band between 70% to 92%, with the exception of period 0 in the **Inconsistent** scenario.

**Examining simulation RMSE**
Combining insights from above, Figure 5 displays root-mean-squared simulation error of methods. Across scenarios and periods, ASCM has the lowest RMSE, at 1.033, while AR-DB has the highest RMSE, at 1.85. Within the DID methods, DD-ES has the lowest RMSE, at 1.52.

**Figure 5: Mean RMSE across simulation iterations**

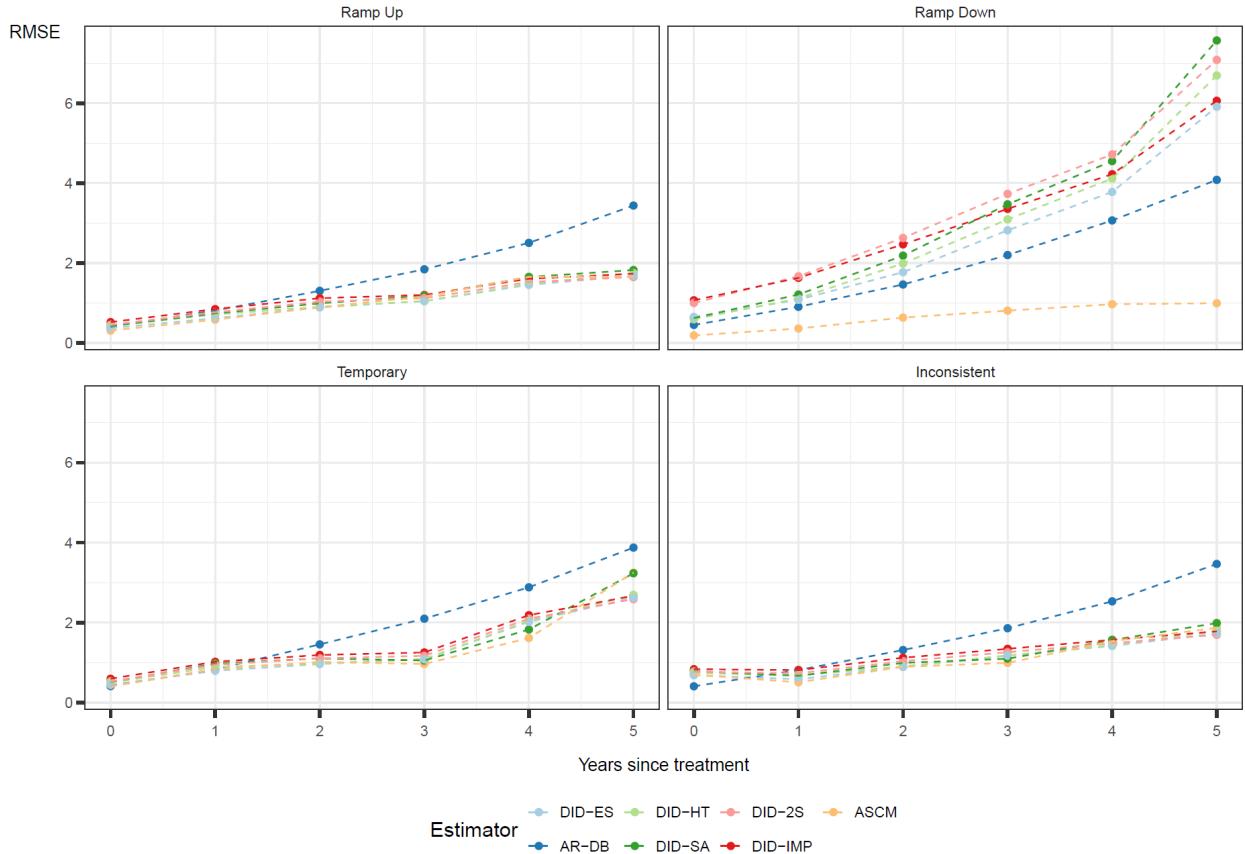

Within the **Ramp down** scenario, ASCM has the lowest average RMSE of 0.65 across time-periods, substantially lower than all other methods (compared to the second lowest, AR-DB, with RMSE = 2.02). However, DID-ES has the lowest average RMSE in all other scenarios (RMSE = 1.01, 1.32, 1.078 in the **Ramp down**, **Temporary**, **Inconsistent** respectively) but is followed closely by ASCM (1.04, 1.34, 1.08). The largest discrepancy in method performance occurs in the **Ramp down** scenario, where DID-2S has an RMSE of 3.47, 433% larger than the RMSE of ASCM. Furthermore, DID methods have particularly poor performance when true effects are diminishing in size, as exhibited in the **Ramp down** scenario. Finally, we find the smallest discrepancy between methods in the **Temporary** scenario, with AR-DB having a RMSE of 1.92, 43% larger than the RMSE of DID-ES.

## DISCUSSION

This study evaluated how different panel data estimators perform when estimating time-varying policy effects under a range of realistic policy scenarios. Our findings highlight the substantial variability in bias, variance, coverage, and root mean squared error (RMSE) across methods, underscoring the importance of selecting an appropriate approach based on the expected policy dynamics. Notably, our results indicate that no single method uniformly outperforms the others across all scenarios, reinforcing the need for researchers to carefully consider methodological trade-offs when conducting state policy evaluations.

A key contribution of this study is demonstrating that augmented synthetic control (ASCM) consistently recovers policy effects with lower bias compared to other approaches, particularly in scenarios where policy effects diminish over time. This finding is especially relevant given that many real-world policies, such as vaccination mandates or substance use interventions, often exhibit declining effectiveness over time. However, ASCM's improved bias performance comes at the cost of increased variance, reflecting the well-known bias-variance tradeoff in statistical estimation.[24,25]

Among DID approaches, DID-2S exhibited the weakest performance, while DID-SA emerged as the most reliable, offering lower bias and reasonable coverage at the expense of slightly higher standard errors. Notably, all DID-based methods struggled in the ramp-down scenario, where policy effects decrease over time. This limitation has important implications for policy research, as failing to account for time-varying dynamics could lead to overestimating the long-term effectiveness of policies.

Our findings also emphasize that researchers should critically assess whether a policy's effects are likely to increase, decrease, or vary unpredictably over time. In cases where effects are expected to diminish, ASCM is preferable over DID methods, which exhibited higher bias and lower coverage in such settings. The tendency of DID methods to produce statistically significant estimates even when policy effects are declining could result in misleading conclusions, potentially influencing policymakers to continue ineffective policies.

Ultimately, the choice of estimation method should be driven by the policy context and the relative costs of bias versus variance. In high-stakes evaluations, such as assessing the effectiveness of opioid-related policies, minimizing bias may be the priority. In exploratory analyses, where the primary goal is to detect potential policy impacts, methods with lower variance may be preferable. Given the methodological heterogeneity in performance, researchers should consider employing multiple approaches and reassessing methods as additional data become available over time.

The study also highlights the broader challenge of estimating dynamic policy effects when treatment effects do not follow a simple linear trajectory. While ASCM appears to be the most reliable in capturing complex time-varying effects, it is not without limitations. For example, in the temporary effects scenario, ASCM exhibited periods of higher bias, suggesting that researchers should be cautious when interpreting results for non-monotonic policy effects.

Despite its contributions, this study has limitations. The simulated scenarios, while designed to reflect realistic policy dynamics, do not capture the full complexity of real-world data, including confounding and selection biases.[8,9] Future research should explore how these methods perform under such conditions and assess their robustness when key assumptions are violated. Additionally, expanding simulations to incorporate larger sample sizes or alternative policy effect structures may provide further insights into the generalizability of these findings.

In conclusion, this study provides clear guidance on the strengths and weaknesses of panel data estimators for estimating time-varying policy effects. By demonstrating that method performance depends critically on the underlying policy dynamics, we highlight the need for careful methodological selection in policy evaluation. Researchers should explicitly consider the expected trajectory of policy effects when choosing estimation methods and remain open to employing multiple approaches to ensure robust and credible policy assessments.

# Supplemental Digital Content

Section 1. Model Estimates and Standard Errors

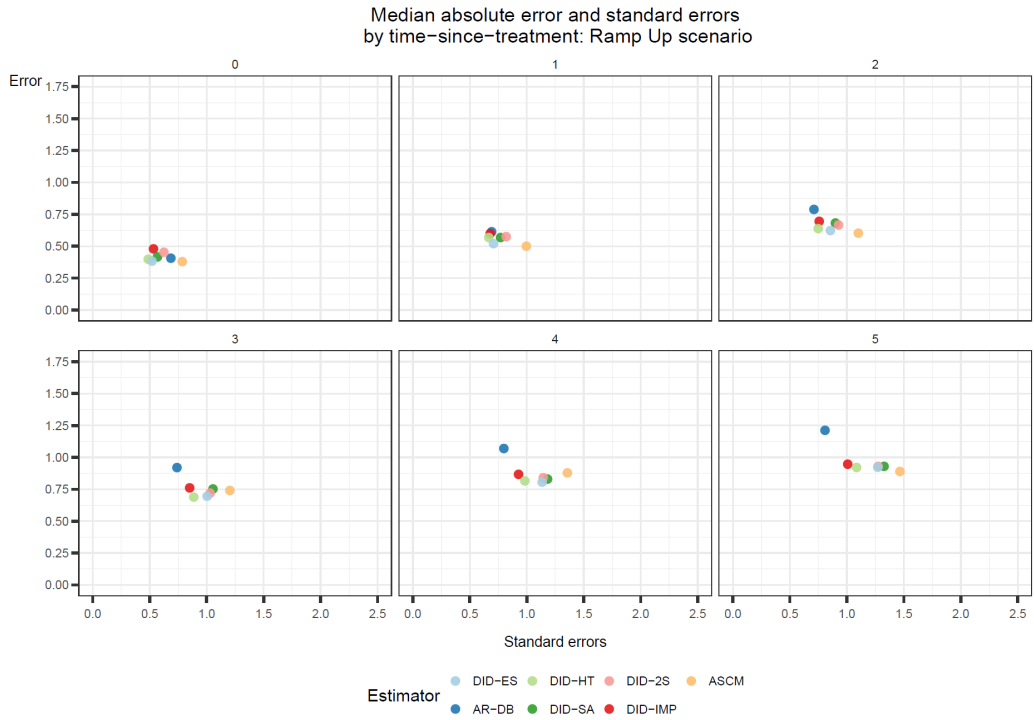

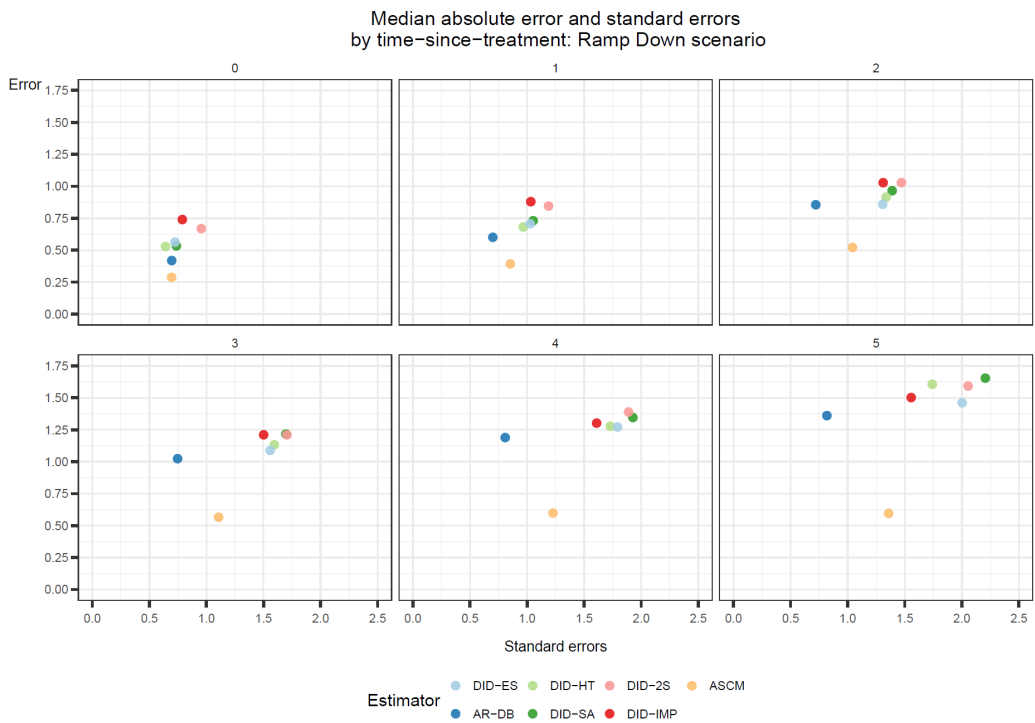

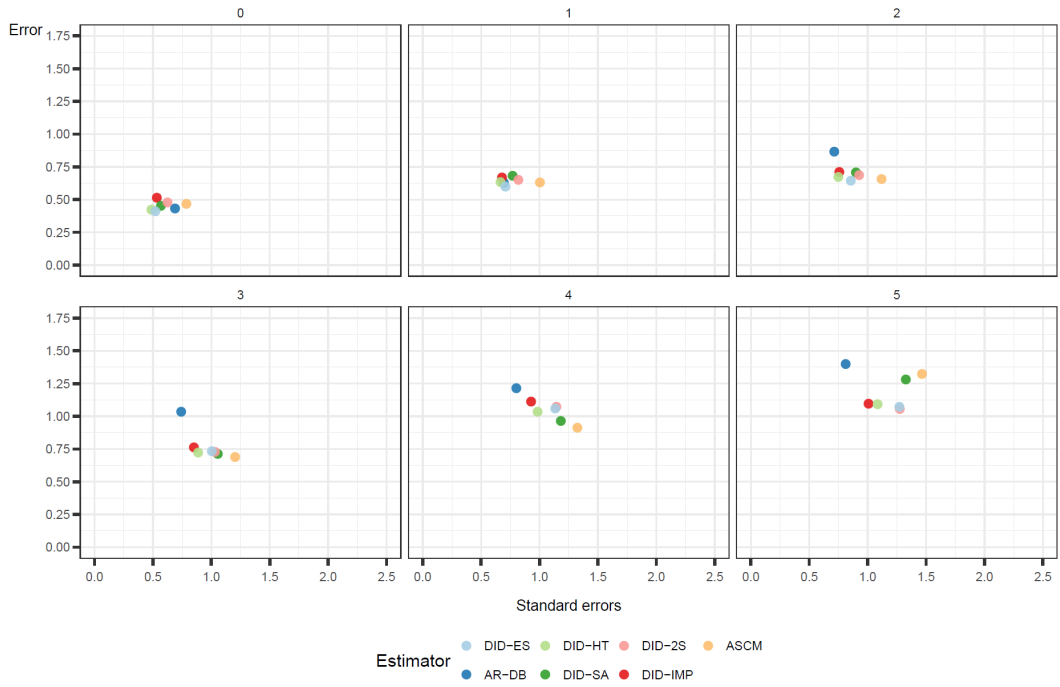
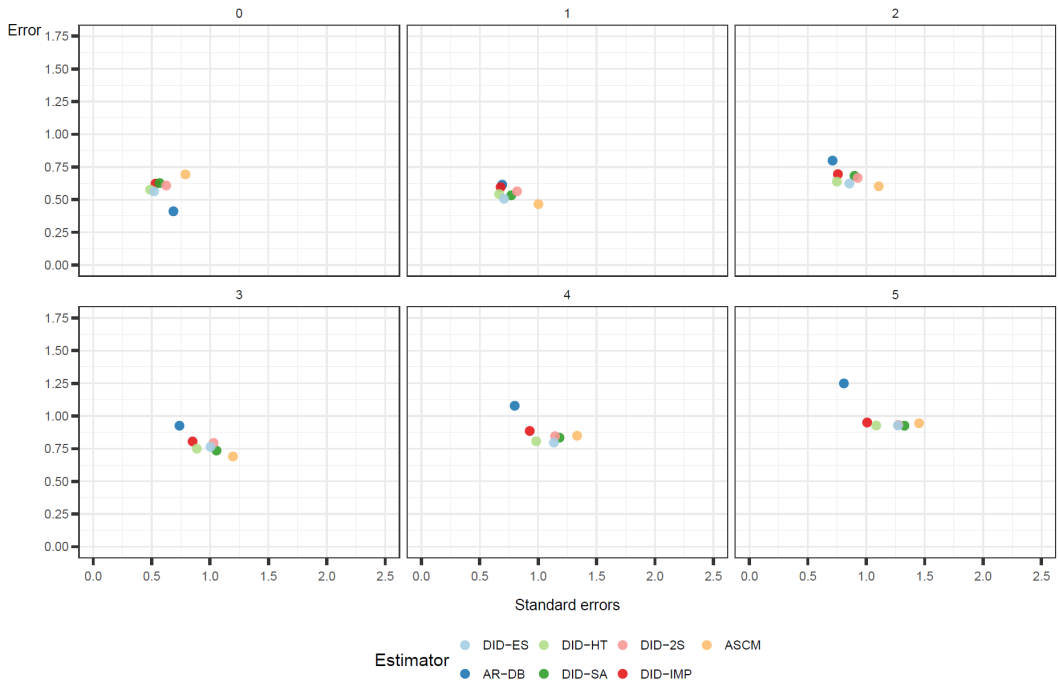

## Section 2. Results Varying Number of Treated Units

Treated: 5 Units

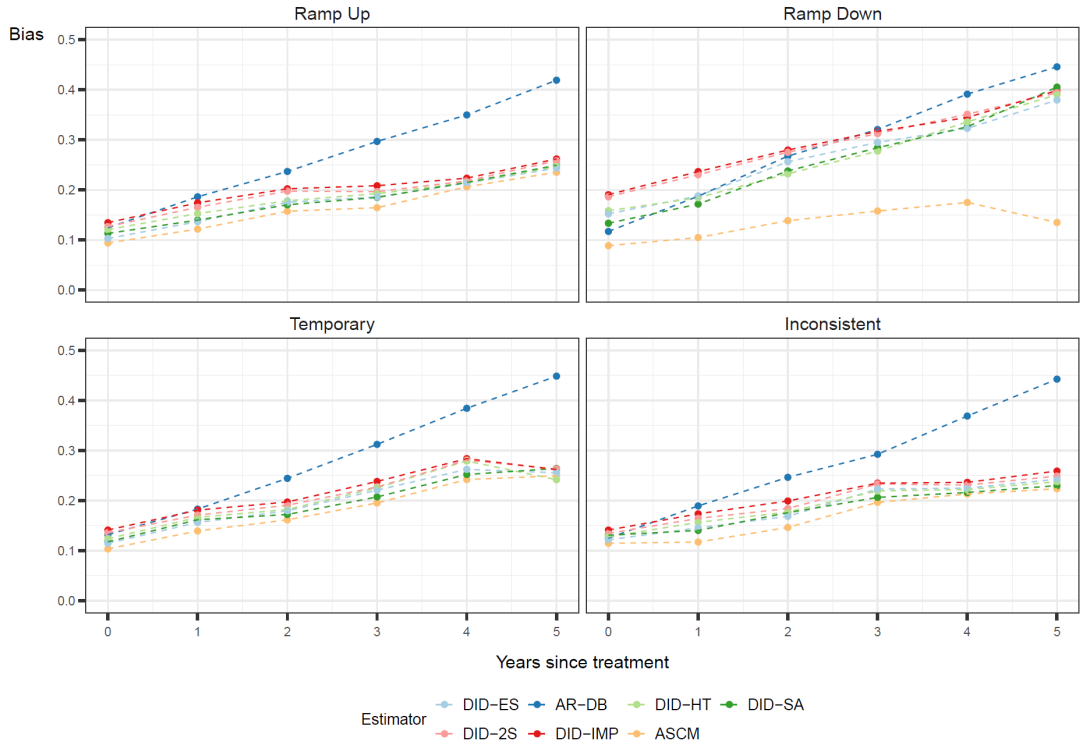

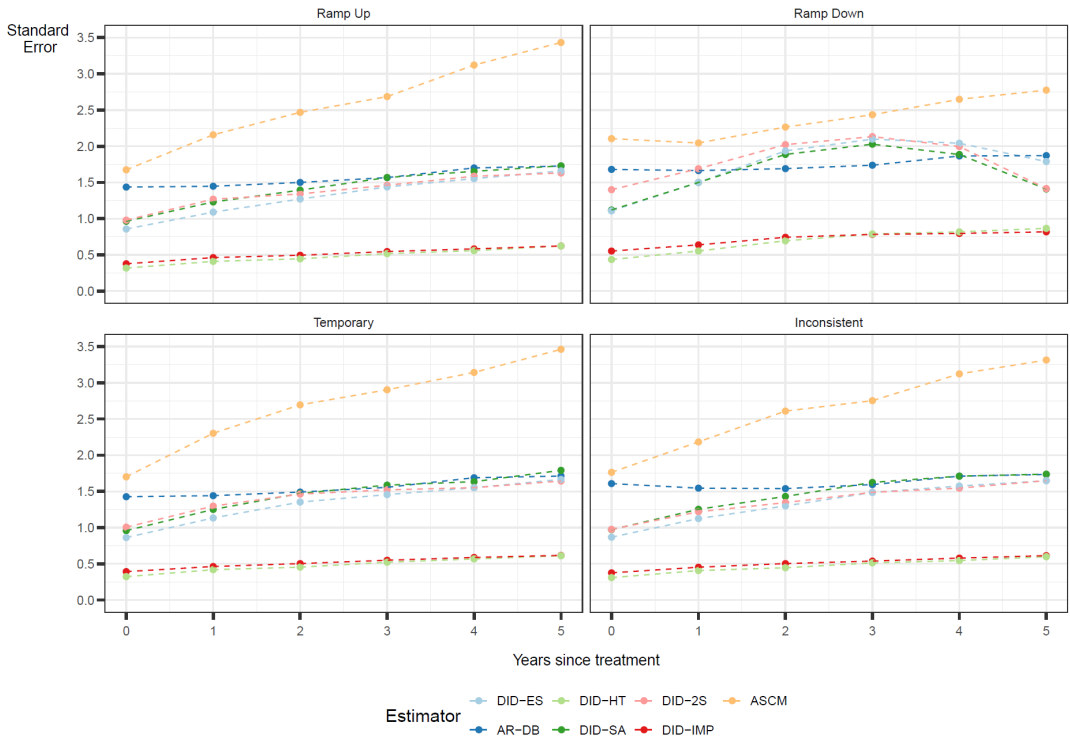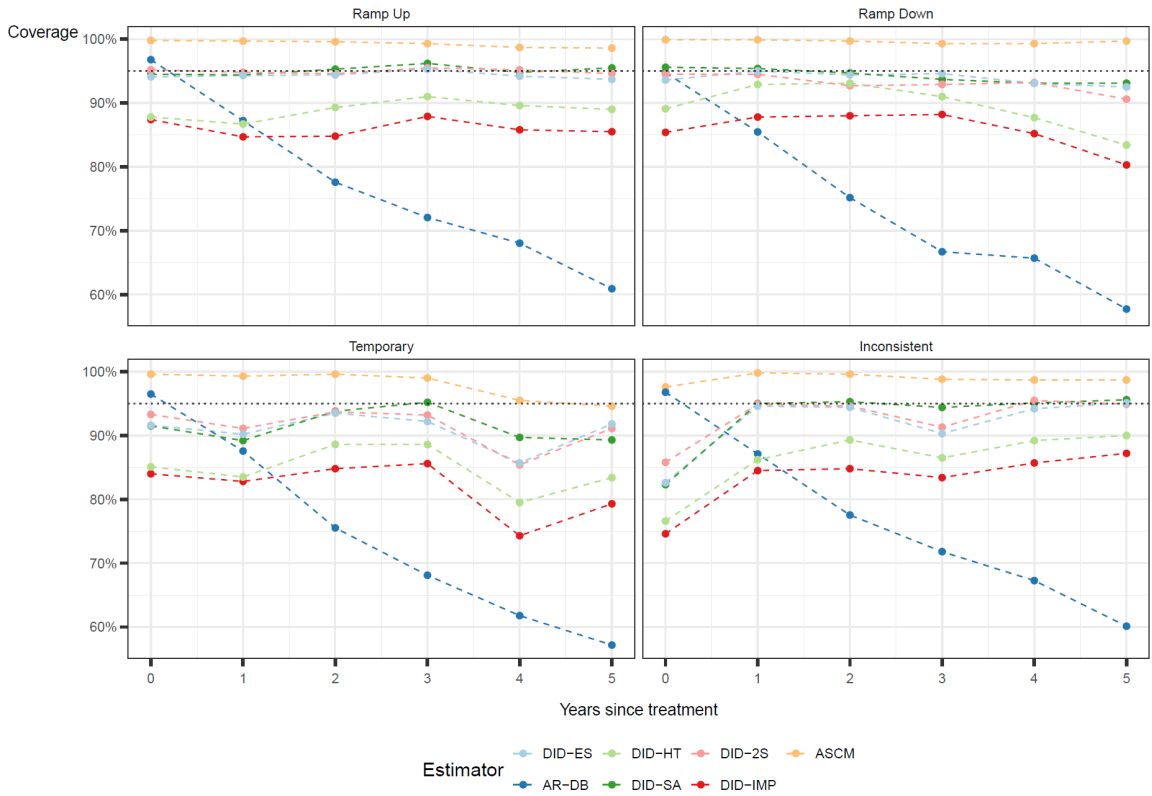

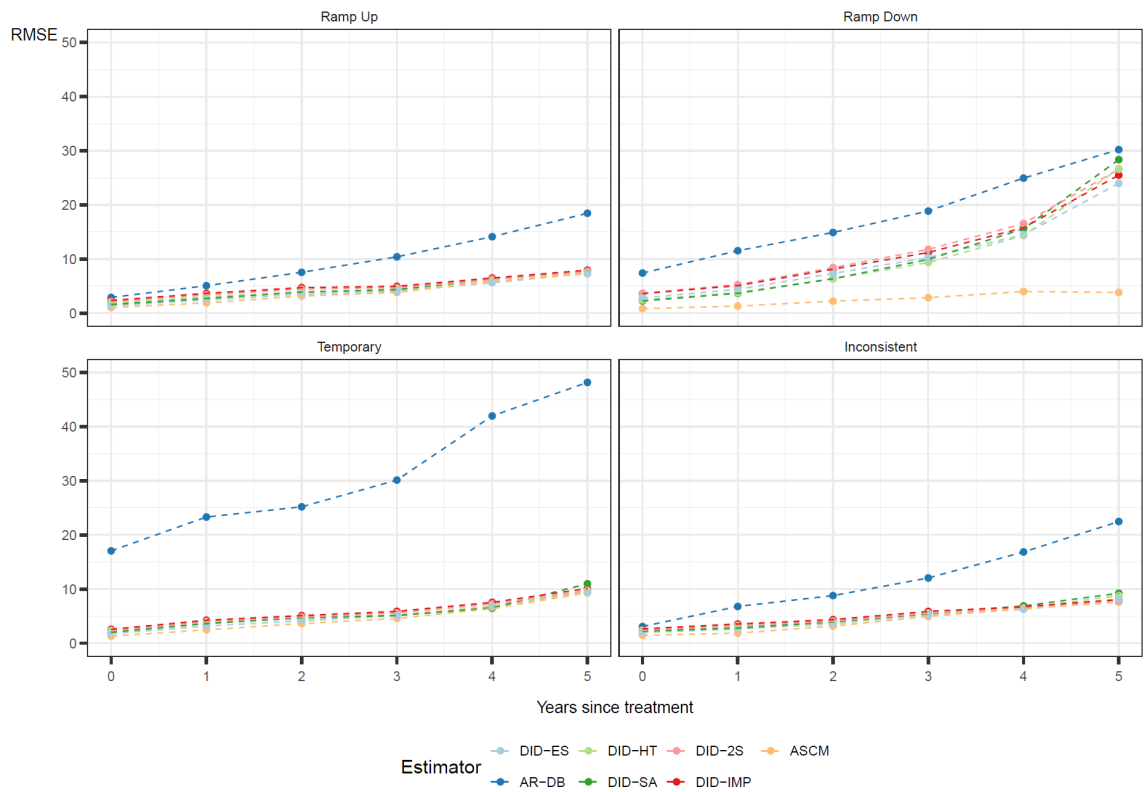

Treated: 45 Units

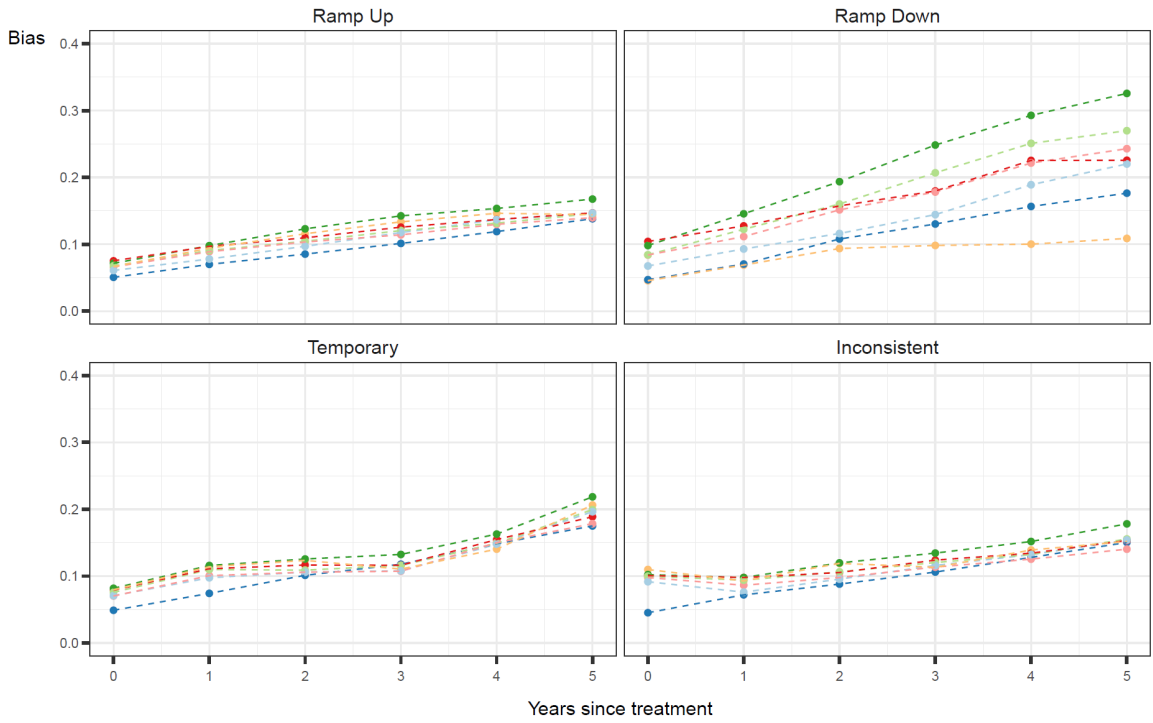

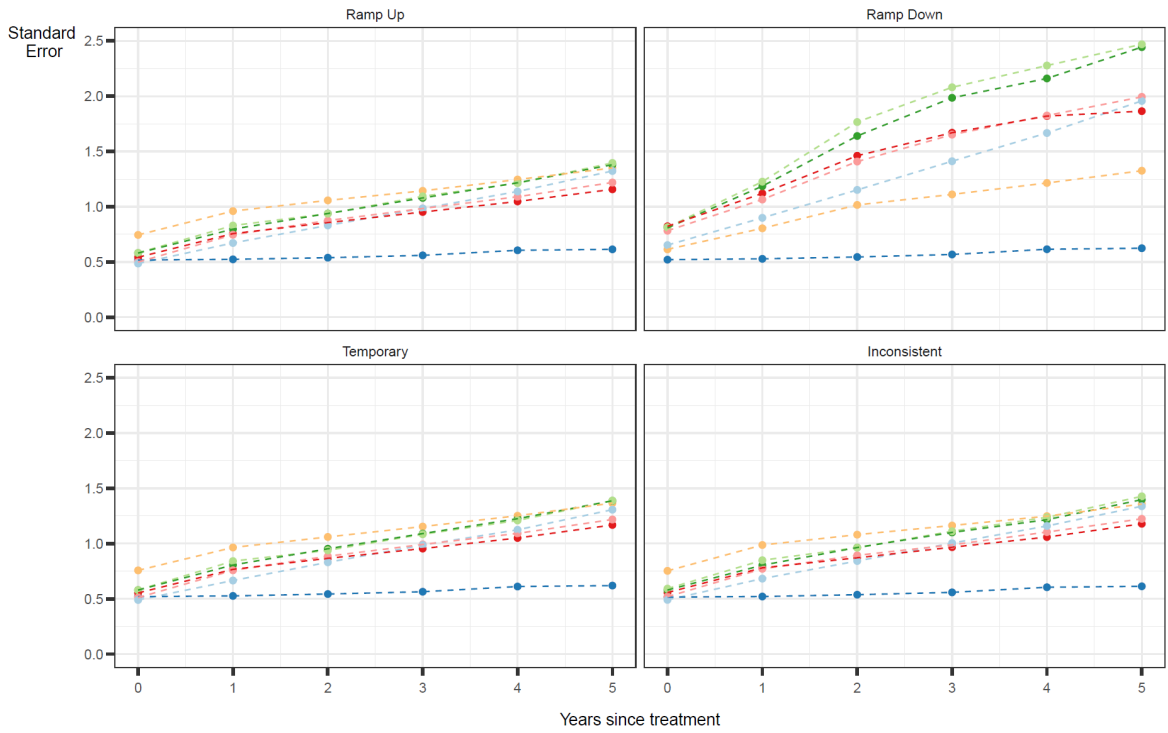

**Section 3. Results with Additional Sample Size**

Methods

We developed 510 synthetic state-years using the synthpop package in R.[26] Synthetic state-year observations using a random forest as the synthesizing method based on the conditional distribution of states, years, crude overdose mortality rates, and unemployment rates. Synthetic variables were constructed sequentially, first synthesizing unemployment rate, followed by crude overdose mortality rates. Synthetic variables were then smoothed to reduce stochastic deviations using a smoothing spline. We then visually compared the synthetic empirical densities of each variable by year to ensure they were reasonably comparable to observed empirical densities (figures below).

Finally, we used the same approach as described in the methods section of the main report to generate policy effect simulations. Synthetic states were merged with observed states to create a dataset with 560 state observations across the study period. We generated 1000 simulations across each scenario, simulating 280 treated units.

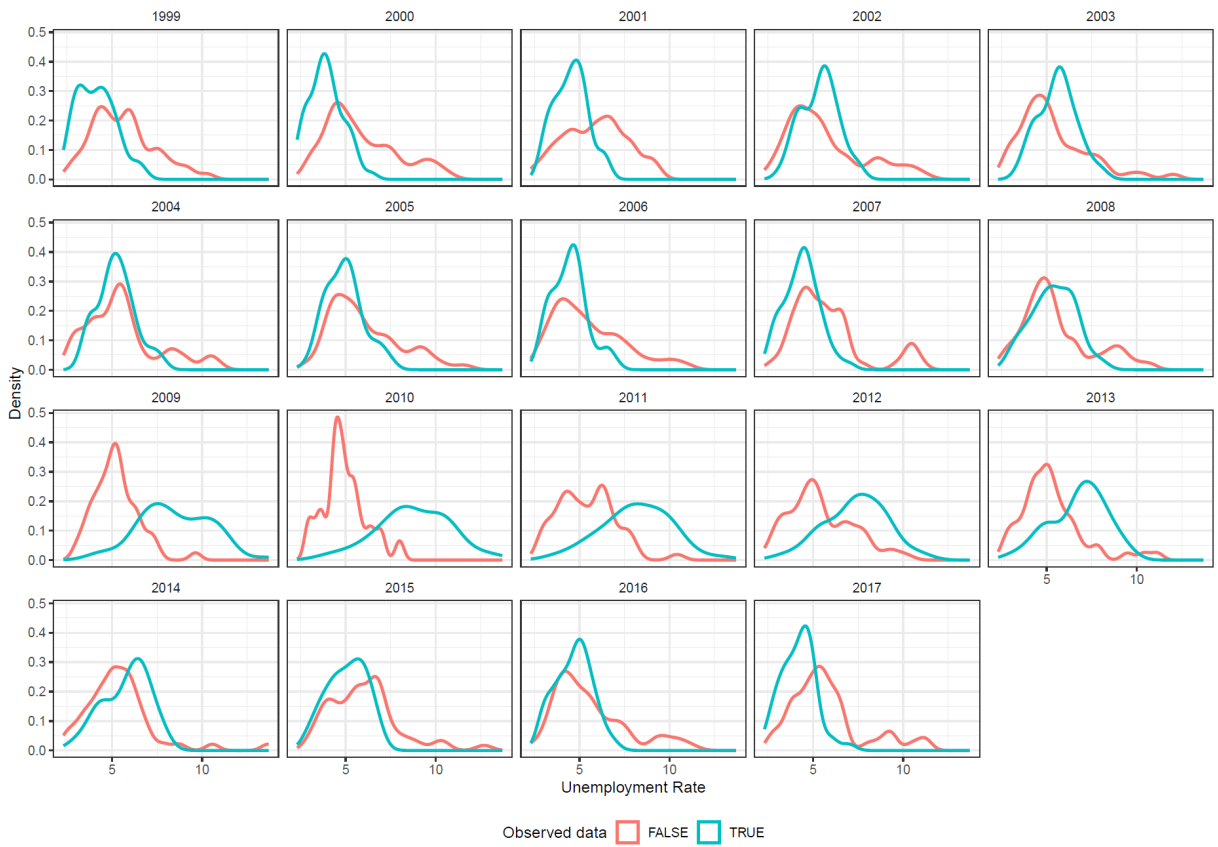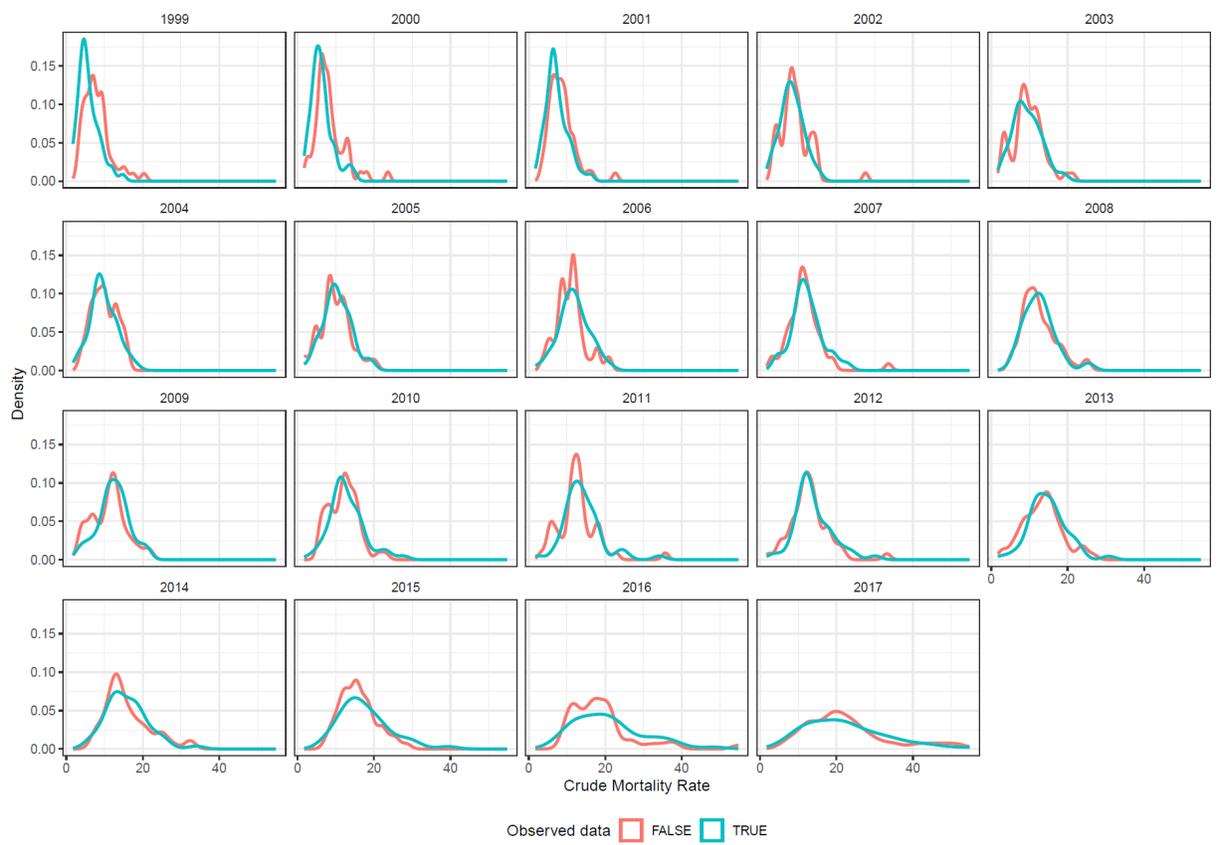

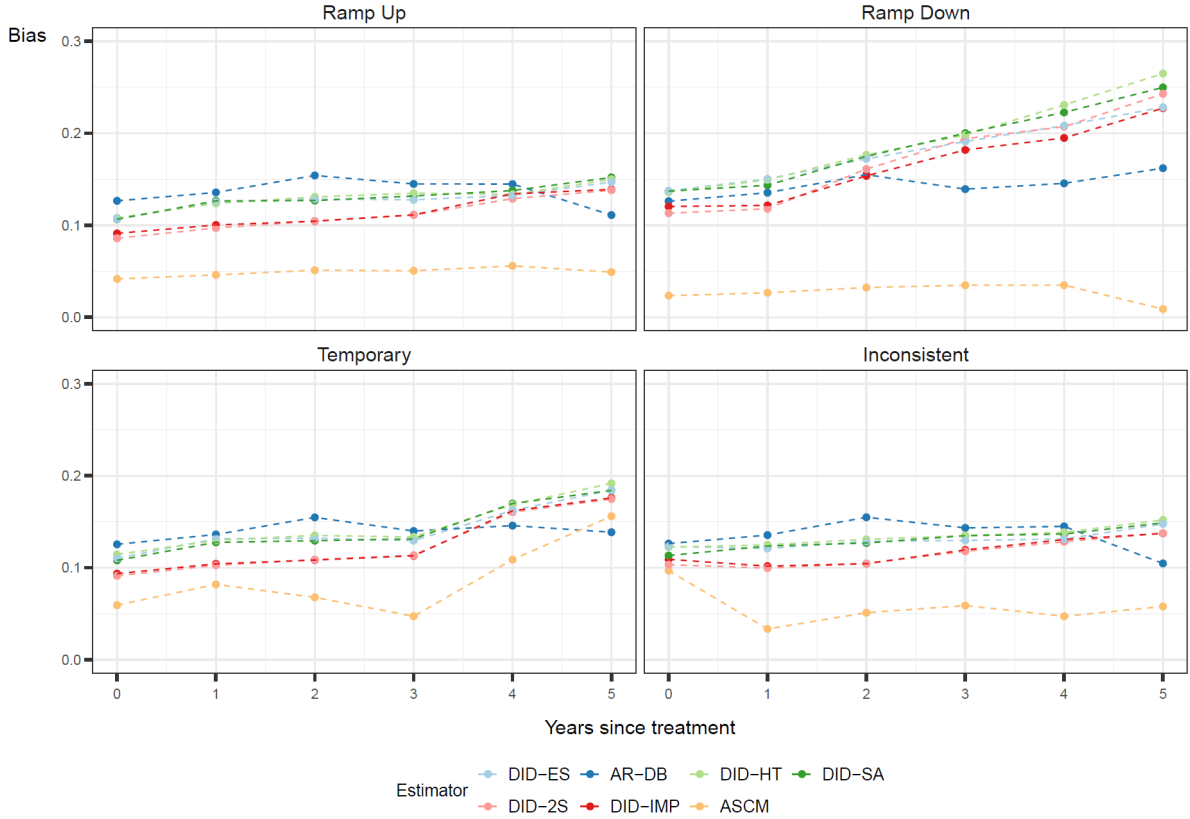

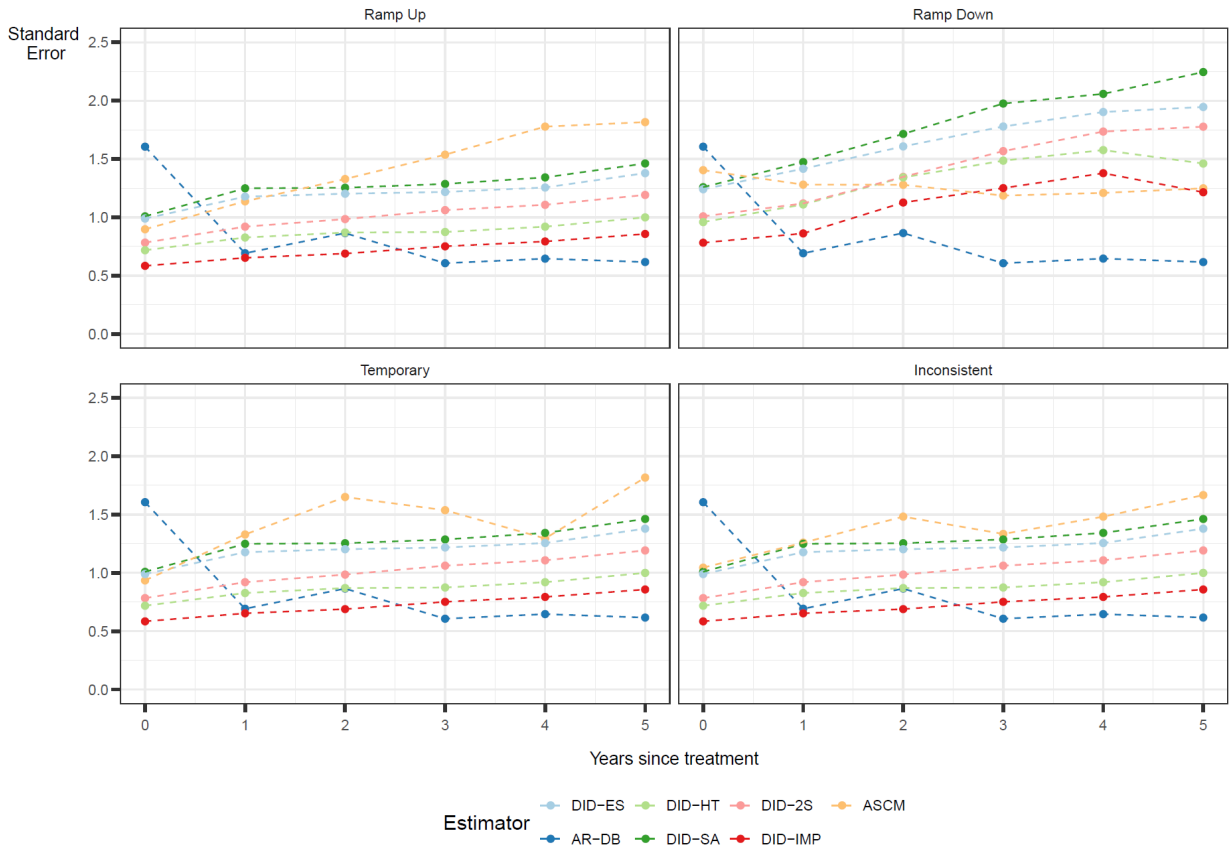

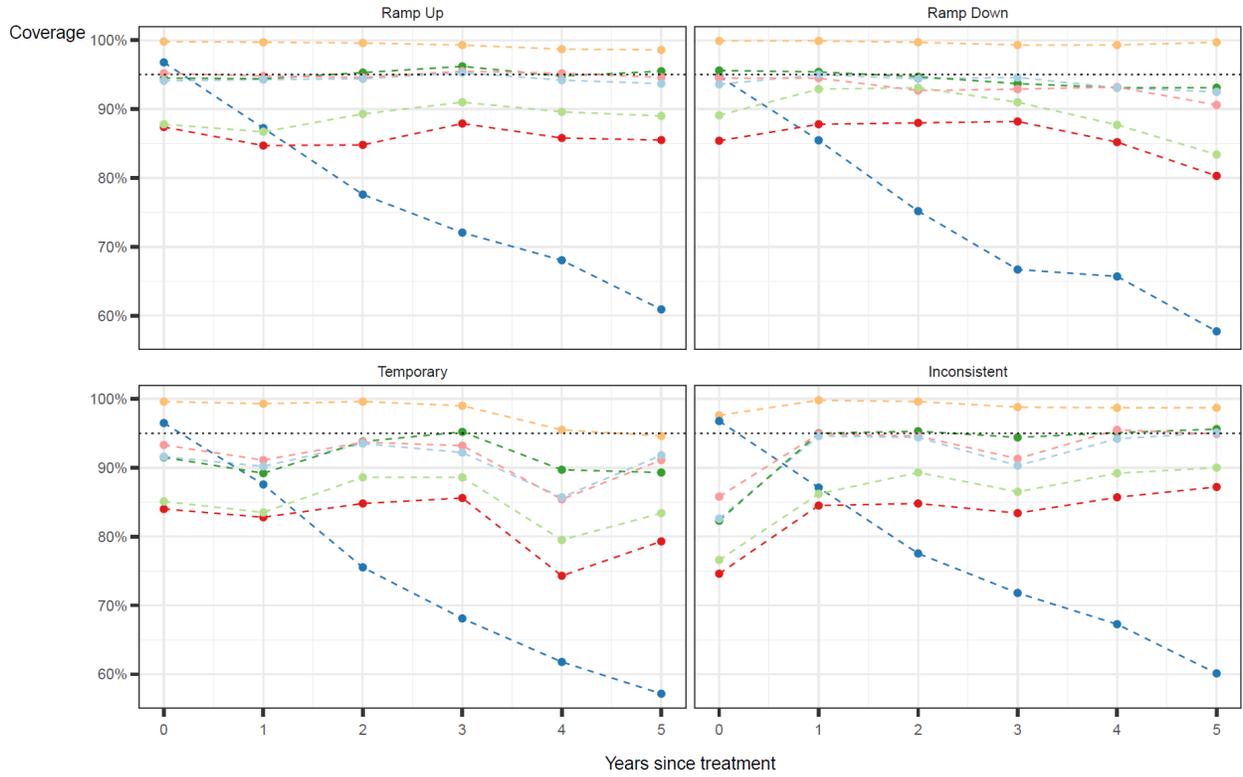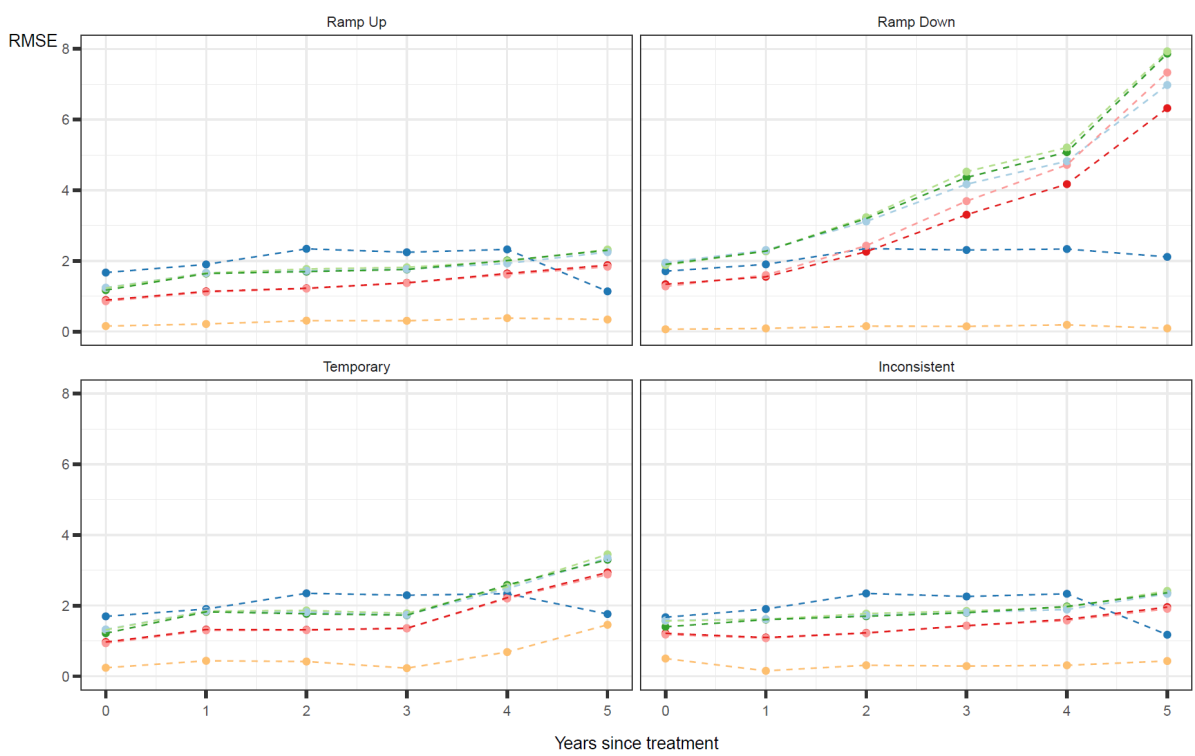